\documentstyle[psfig,graphicx]{mn2e}
\footnotesize
\newdimen\minuswidth    
\setbox0=\hbox{$-$}
\minuswidth=\wd0
\catcode`@=\active
\def@{\kern\minuswidth}
\newdimen\digitwidth    
\setbox0=\hbox{\rm0}
\digitwidth=\wd0
\catcode`!=\active
\def!{\kern\digitwidth}
\normalsize

\title[A Very Large Glitch in PSR~J1806$-$2125]
{A Very Large Glitch in PSR~J1806$-$2125}
\author[G. Hobbs et al.]
{G. Hobbs,$^1$
A. G. Lyne,$^1$
B. C. Joshi,$^1$
M. Kramer,$^1$
I. H. Stairs,$^2$ 
\newauthor
F. Camilo,$^3$
R. N. Manchester,$^4$
N. D'Amico,$^{5,6}$ 
A. Possenti$^5$
and
\newauthor
V. M. Kaspi$^{7,8}$
\\
$^1$ University of Manchester, Jodrell Bank Observatory, Macclesfield,
Cheshire SK11~9DL \\
$^2$ National Radio Astronomy Observatory, PO~Box~2, Green Bank,
WV~24944, USA \\
$^3$ Columbia Astrophysics Laboratory, Columbia University, 550 West
120th Street, New York, NY~10027, USA \\
$^4$ Australia Telescope National Facility, CSIRO, PO~Box~76, Epping
NSW~1710, Australia \\
$^5$ Osservatorio Astronomico di Bologna, via Ranzani 1, 40127~Bologna,
Italy \\
$^6$ Cagliari Astronomical Observatory, Loc. Poggio dei Pini, Strada
54, 09012 Capoterra (Ca), Italy \\
$^7$ Massachusetts Institute of Technology, Center for Space Research,
70 Vassar Street, Cambridge, MA~02139, USA \\
$^8$ Physics Department, McGill University, 3600 University Street,
Montreal, Quebec, H3A~2T8, Canada \\
}
\let\leqslant=\leq 
\date{2002 April 16}
\begin{document}

\maketitle
\newcommand{\setthebls}{
}

\setthebls

\begin{abstract}
 PSR~J1806$-$2125 is a pulsar discovered in the Parkes multibeam
 pulsar survey with a rotational period of 0.4~s and a
 characteristic age of 65~kyr.  Between MJDs 51462 and 51894 this
 pulsar underwent an increase in
 rotational frequency of $\Delta \nu/\nu \approx 16 \times
 10^{-6}$.  The magnitude of this glitch is $\sim$2.5 times greater
 than any previously observed in any pulsar and 16 times greater than
 the mean glitch size. This letter gives the parameters of the 
 glitch and compares its properties to previously
 observed events. The existence of such large and rare glitches offers new 
 hope for attempts to observe thermal X-ray emission from the 
 internal heat released following a glitch, and suggests that pulsars which
 previously have not been observed to glitch may do so on long timescales.
\end{abstract}


\section{INTRODUCTION}
The spin-down of pulsars is usually remarkably steady and predictable. However,
timing observations of young pulsars have revealed rotational
irregularities such as timing noise, a noise-like fluctuation in
rotation rate, and glitches which manifest themselves in a sudden spin-up
of the pulsars (e.g.~Lyne, Shemar \& Graham-Smith 2000\nocite{lsg00}). 
Typical increases in rotational frequency during a glitch are
of the order of $\Delta \nu/\nu=10^{-8}$ to $10^{-6}$ which is 
followed by a relaxation process during which the pulsar usually 
returns to its pre-glitch spin-down rate. The time scales for
relaxation vary from hours to years, depending on pulsar and glitch event.

Observing glitches and their relaxation processes provides a unique method for
studying the interior of neutron stars: glitches are thought to be caused
by a sudden transfer of angular momentum from a faster-rotating component
of the superfluid interior to the solid crust of the pulsar. Hence,
monitoring pulsars to detect glitches and to measure their subsequent
relaxation provides a kind of rotational seismology
to probe neutron star interiors (e.g. Pines 1991\nocite{pin91}; 
Lyne 1992\nocite{lyn92a}). 

In this Letter we present the largest glitch event ever observed, with 
a fractional frequency increase almost 2.5 times 
larger than the previously known largest glitch \cite{wmp+00} 
and 16 times greater than the mean glitch size. This glitch was
observed in PSR~J1806$-$2125, a pulsar which was discovered in the 
Parkes multibeam survey \cite{mlc+01}.  This survey has 
discovered numerous young pulsars, which are
subsequently monitored using the Parkes 64-m or the Jodrell Bank 76-m
radio telescopes.  The discovery of PSR~J1806$-$2125 is reported in
Morris et al. (2002)\nocite{mhl+02}. The pulsar has a rotational 
period of $P=0.4$~s and a characteristic age of 65~kyr.  

\section{Observations}
 Following its discovery, PSR~J1806$-$2125 was observed 37 times using the 76-m 
 Lovell radio telescope at Jodrell Bank Observatory between 
 January 1998 and October 1999 (MJDs~50820 and 51462), and again 
 from January 2002 (MJD~52298).   The observing system is 
 described in Morris et al. (2002)\nocite{mhl+02}. In brief, the two 
 hands of circular polarisation at a frequency near 1400~MHz are fed 
 through a multichannel filterbank and digitised.  The data are
 dedispersed and folded on-line according to the pulsar's dispersion
 measure and topocentric period.  The folded pulse profiles for each
 polarisation are subsequently combined to produce the total intensity.
 Pulse times-of-arrival (TOAs) are determined by cross-correlating the
 profile with a template of high signal-to-noise ratio.  During the
 upgrade period of the Lovell telescope, timing observations
 are continuing using the central beam of the 13-beam system installed on 
 the 64-m Parkes radio telescope \cite{mlc+01}.  Between December 2000
 and January 2002 (MJDs~51894 and 52257), the pulsar was observed 14
 times with a typical integration time of 240\,s. 

\section{Results}
 \begin{table}\begin{center}
  \caption{Observed and derived parameters for PSR~J1806$-$2125.  
           The characteristic age is calculated as $P/(2\dot{P})$,
           the surface magnetic dipole field strength 
           as $3.2\times10^{19}(P\dot{P})^{1/2}$~Gauss and the rate of
	   loss of rotational energy as $4\pi^2 I\dot{P}P^{-3}$ where
           a neutron star with moment of inertia $I = 10^{45}$\,g\,cm$^2$ is assumed.}
  \begin{tabular}{lll} \hline
   Right ascension (J2000)  & 18$^{\rm h}$06$^{\rm m}$19$.^{\rm s}$59(8) \\
   Declination (J2000)      & $-$21\degr25\arcmin40\arcsec(24) \\
   Period (s)         & 0.48178844377(6) \\
   Period derivative (10$^{-15}$) & 117.295(14) \\
   Frequency, $\nu$ (Hz)     & 2.0755998051(2)  \\
   Frequency derivative, $\dot \nu$ (10$^{-15}$ s$^{-2}$) & $-$505.32(6)\\
   Period/frequency epoch (MJD) & 51062.8 \\
   Dispersion measure (cm$^{-3}$pc)& 750(3)\\ \\
   rms timing residual (ms) & 3.5 \\
   Epoch range (MJD) & 50820--51305 \\ 
   Flux Density at 1400~MHz (mJy) & 1.1(2) \\ \\

   Characteristic age (kyr) & 65\\
   Surface magnetic field (10$^{12}$ G)& 7.8\\ 
   Rate of loss of rotational energy (erg\,s$^{-1}$) & 4.1$\times$10$^{34}$  \\ \hline
  \end{tabular}
  \label{tb:posnparam}
 \end{center}\end{table}

\begin{figure}
  \centerline{\includegraphics[width=80mm]{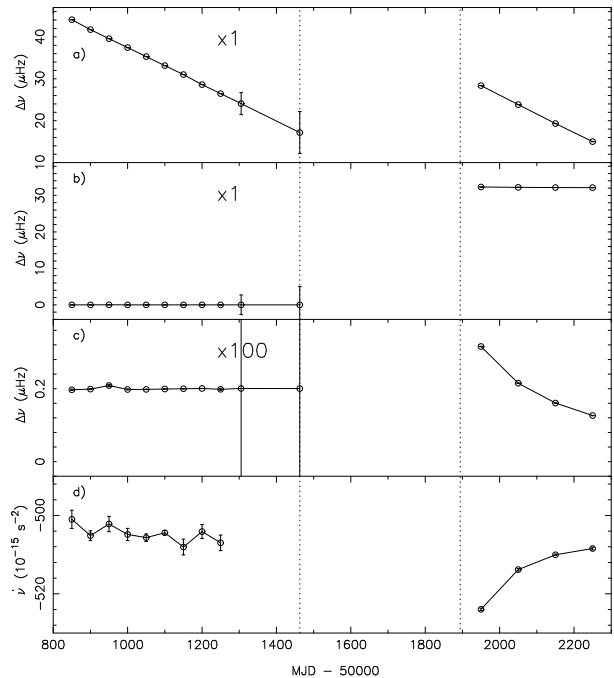}}
  \caption{a) The rotational frequency, $\nu$, offset by 2.075565~Hz,
           between MJDs 50850 and
           52250, b) the same after subtracting the ephemeris in Table
           \ref{tb:posnparam}, c) to observe structure in the post-fit
           data an arbitrary offset has been added and the scale
           increased by a factor of 100 and d) the frequency derivative,
           $\dot{\nu}$. To obtain these plots, fits were made to
           pulse arrival times to obtain frequency and frequency
           derivatives assuming the position and
           dispersion measure given by the ephemeris in Table \ref{tb:posnparam}.
           The dashed lines indicate the last Jodrell Bank observation and
           first Parkes observation before and after the glitch.  In
           most cases the uncertainties are smaller than the size of
           the symbol.}
  \label{fg:glitch}
 \end{figure}

 The rotational, positional and derived parameters for PSR~J1806$-$2125
 are given in Table \ref{tb:posnparam}.  These parameters are
 obtained by model-fitting the Jodrell Bank pulse TOAs 
 using \textsc{tempo}\footnote{See
 http://pulsar.princeton.edu/tempo.}.  All uncertainties are twice the 
 standard \textsc{tempo} values.
 The rotational frequency of the pulsar increased between October 1999
 and December 2000, indicating that a glitch had occurred.
 The nature of the event is summarised in Figure \ref{fg:glitch}.  
 The rotational frequencies versus time 
 are plotted in Figure \ref{fg:glitch}a. It is clear that sometime during the
 430~day gap between observations, approximately 700 days of normal spin-down
 were reversed.  Assuming, as argued later, that this occurred in a single event, the step
 change in rotation rate was $\Delta \nu/\nu \approx
 16\times10^{-6}$. The frequencies for the final two pre-glitch Jodrell Bank observations
 were obtained by determining the shift in the pulse profile across
 the integration time of $\sim$20~minutes.  All other frequencies were obtained from 
 a least-squares fit of a timing model to 4--8 adjacent TOAs, 
 keeping positional parameters and the frequency
 derivative fixed with the epoch set to the mid-point of the TOAs.
 The effect of subtracting the pre-glitch 
 rotational frequency and frequency derivative, determined from 
 the Jodrell Bank data, is shown in Figure \ref{fg:glitch}b.   
 To view structure in the post-glitch residuals, an offset is
 subtracted from the post-fit data and the scale expanded by a factor
 of 100 (Figure \ref{fg:glitch}c).  
 The post-glitch rotational frequency decays with time over a few hundred
 days.  The changing frequency derivative is
 shown in Figure \ref{fg:glitch}d.  The last observation
 prior to the glitch was obtained at Jodrell Bank on MJD~51462 and the
 first observation after the glitch was obtained at Parkes on MJD~51894 (indicated by
 dotted lines in Figure \ref{fg:glitch}).  Unfortunately,
 the large interval between the Jodrell Bank and Parkes observations
 prevents extrapolation of the pre- and post- glitch pulse 
 ephemerides without pulse period ambiguities.  We can therefore
 only deduce that the glitch occurred sometime between the two dates.
 
 \begin{figure*}
  \centerline{\includegraphics[width=30mm,angle=-90]{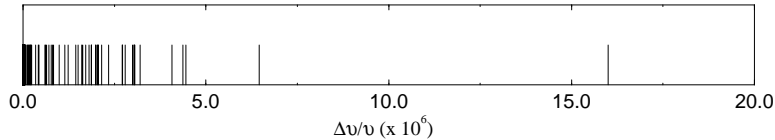}}
  \caption{Fractional change in rotational frequency 
          $(\Delta \nu/\nu \times 10^6)$ for the 97 glitches published
          in Joshi et al. (2002) and the glitch described here.}
  \label{fg:hist}
 \end{figure*}

 Table \ref{tb:param} gives the pre-glitch value of the frequency,
 $\nu$, and its first and second derivatives extrapolated to the epoch of the
 first observation after the glitch (MJD~51894).  This table also
 contains the analogous post-glitch parameters given for the same
 epoch and the instantaneous changes at the glitch. For the pre-glitch
 data, $\ddot{\nu}$ reflects timing noise, however in the post-glitch
 data, $\ddot{\nu}$ describes both the timing noise and the recovery
 from the glitch. No other glitch
 events are visible in the data limiting the magnitudes of any
 glitches to less than $\Delta \nu/\nu \approx 
 10^{-9}$.   In some cases, glitches may be
 confused with timing noise particularly when there are large gaps between 
 observations.  The timing residuals of PSR~J1806$-$2125
 show small amounts of timing noise; the pre- and post-glitch residuals
 (Table~\ref{tb:param}) have root-mean-square values of 3.4 and 8.2~ms 
 respectively.  As the pre- and post-glitch data sets
 are only  $\sim$1 year in length, these fits may 
 under-estimate the amount of timing noise which may have been absorbed in the
 rotational parameters or position.  In particular, the position given
 in Table \ref{tb:posnparam} may be in error by more than the quoted
 formal uncertainty.  Even if PSR~J1806$-$2125 had similar timing
 noise properties to PSR~B1951$+$32, which is known to exhibit extreme
 amounts of timing noise \cite{flsb94}, its apparent fractional
 frequency change through the data would still be less than $\sim
 10^{-8}$. This would not affect the main parameters of the glitch given in Table
 \ref{tb:param}.


 \begin{table*} \begin{center}	
 \caption{Pre-glitch, post-glitch and glitch parameters extrapolated
          to MJD~51894.  The root-mean-square timing residuals for
          each fit are given in the final column.  The quoted errors
          in parentheses are twice the formal errors in the last quoted digit
          and are obtained from a least squares fit of a timing model 
          to the TOAs.}
  \begin{tabular}{lcllll}\hline
	      & Fit Interval & $\nu$  & $\dot{\nu}$  & $\ddot{\nu}$ & Residual \\
	      & (MJD)        & (Hz)  & (10$^{-15}$s$^{-2}$) & (10$^{-24}$s$^{-3}$) & (ms) \\ \hline
  Pre-glitch  & 50820--51305  & 2.07556349(2) & $-$505.9(8)   & $-$77(10) & 3.4 \\ 
  Post-glitch & 51894--52257  & 2.075595904(10) & $-$523.0(16) & $+$620(120) & 8.2\\ 
  Glitch increment  &        & 0.00003241(2)   & !$-$17(2)    &$+$697(120)   &  -- \\ \hline 
  \end{tabular}
 \label{tb:param}
 \end{center} \end{table*}

\section{Discussion} 
 Pulsar glitches seem to be divided into two sizes: most are large
 with fractional frequency increases of $\sim$10$^{-6}$ while the
 smaller events are in the range $10^{-7}$ to $10^{-9}$.  
 The event described in this letter is large --- the fractional
 frequency increase is more than twice 
 that of the largest previously known glitch (Figure \ref{fg:hist}). 
 It is not possible to rule out the possibility 
 that multiple glitches occurred during the gap of observations between 
 MJDs 51462 and 51894. This large gap, due to an administration error,
 is an unfortunate feature of our data and highlights the
 importance of making regular timing observations of young pulsars.
 However, pulsars with large glitches tend to show larger intervals between glitches 
 (Lyne et al. 2000) \nocite{lsg00} and no glitch occurred during the 600 days of well sampled
 observations at Jodrell Bank. Assuming that the 
 integrated effect of the glitches is to reverse 1.7 per cent of the 
 star's slow-down (Lyne et al. 2000)\nocite{lsg00} then such giant-sized glitches can only
 occur every $\sim$120~years.

\begin{figure}
  \centerline{\includegraphics[width=80mm,angle=-90]{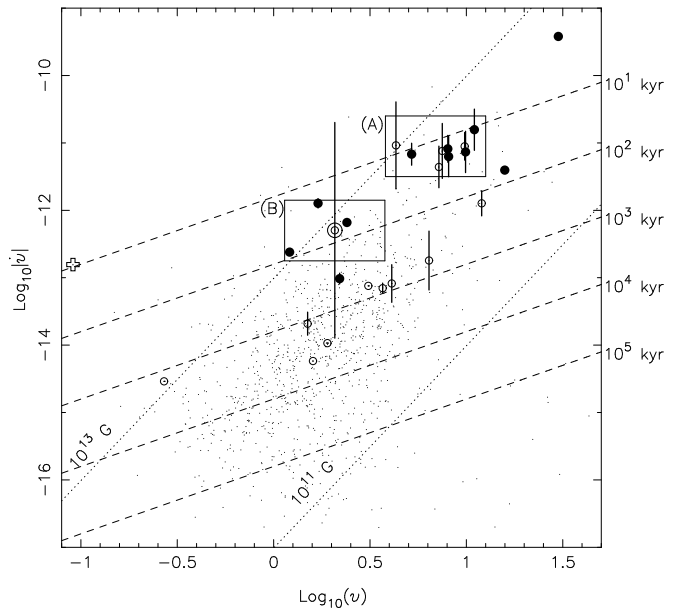}}
  \caption{Pulsars that have glitched, shown on a plot of magnitude of
           frequency derivative versus rotational frequency.
           PSR~1806$-$2125 is indicated by a double circle.  Pulsars
           that have glitched multiple times with a mean time 
           between glitches less than 5
           years are shown as solid circles.  Open circles indicate 
           pulsars that glitch less regularly. The size
           of the vertical lines reflects the mean size of
           glitches.  The cross
           positions the anomalous X-ray pulsar
           1RXS~1708$-$4009 which has also glitched 
           (Kaspi, Lackey \& Chakrabarty 2000). Lines of
           constant characteristic age are shown as dashed lines and
           constant magnetic field as dotted lines.}
  \label{fg:vvdot}
 \end{figure}

  The mean size of
  pulsar glitches on a plot of the magnitude of frequency derivative 
  versus rotational frequency is shown on Figure
  \ref{fg:vvdot}.\nocite{klc00}\nocite{jlk+02}
  Clearly, glitches occur predominantly in young pulsars located in
  the top right of the diagram.  However, further trends are more
  difficult to identify.  Region A on this diagram contains nine 
  glitching pulsars with similar characteristic ages of $\sim$10~kyr
  and five pulsars with no  history of glitching. The surface magnetic
  field strengths for the pulsars in this region 
  range from $10^{12}$ to $10^{13}$~Gauss.  
  Amongst 34 pulsars in an equivalent region centred on 
  PSR~J1806$-$2125 (region B on Figure \ref{fg:vvdot}),
  only four have been observed to glitch.  PSR~B2334$+$61, a pulsar
  which lies just above PSR~J1806$-$2125 in this region, has
  been observed at Jodrell
  Bank for 15 years. Although the timing residuals for this pulsar show large
  amounts of timing noise, no glitch has been observed; any large
  glitch would easily be observable.  This pulsar could therefore
  glitch on a possibly similar 100~year timescale.  However, PSR~B1737$-$30 is
  also in this region and glitches regularly --- 
  nine small glitches over 8.5 years with a mean fractional frequency
  increase of 221$\times$10$^{-9}$ have been reported in Lyne et
  al. (2000) and Krawczyk et
  al. (2002)\nocite{lsg00}\nocite{klg+02}.  
  This suggests that the size of and time between glitches depends upon
  more than the pulsar's position in the frequency--frequency
  derivative diagram.
  
 Large glitch events are clearly very rare and
 will only be detected in pulsars which have been observed for many
 years.  This suggests that many pulsars which
 have not been observed to glitch may do so on longer 
 timescales.  The existence of such large glitches offers new hope for
 attempts to observe thermal X-ray emission from the 
 internal heat released following a
 glitch. Such a large event in a nearby pulsar would
 provide an opportunity for detecting such thermal emission and hence
 constrain the neutron star structure \cite{tc01}.  

 The amplitude of the transient seen in Figure \ref{fg:glitch}c is
 less than 1 per cent of the step change in frequency and is consistent
 with results presented in Lyne et al. (2000).\nocite{lsg00}
 The amplitude could, of course, be much larger if the glitch occurred
 near the beginning of the gap of observations.  
 The moderately large fractional change in frequency derivative ($\Delta
 \dot{\nu}=-17\times10^{-15}$~s$^{-2}$ or
 $\Delta \dot{\nu}/\dot{\nu}=34\times10^{-3}$) suggests that the pulsar's 
 effective moment of inertia was reduced during the glitch by 3 per
 cent.  This reduction is transitory; after $\sim$300 days, $\dot{\nu}$
 approximates its pre-glitch value.

 This is the first observation of such an extreme glitch
 event. Although such large events must be rare, the great number of
 pulsars now known due to the highly successful Parkes multibeam
 survey increases the chance of studying such events in detail.
 Information obtained in these studies will provide unique constraints
 on theories of the internal structure of neutron stars and the
 mechanism which reduces a surprisingly large fraction of the
 effective moment of inertia of these massive cosmic fly-wheels.

\section*{Acknowledgments}
GH acknowledges the receipt of post-graduate studentships from 
the UK Particle Physics and Astronomy
Research Council. IHS received support from the Jansky
postdoctoral fellowship.  FC is supported by NASA grant
NAG~5-9095.  VMK is an Alfred P. Sloan Research Fellow and was
supported in part by a US National Science Foundation (NSF) Career
Award (AST-9875897) and by a Natural Sciences and Engineering Research
Council of Canada grant (RGPIN 228738-00). The Parkes radio telescope
is part of the Australia Telescope which is funded by the Commonwealth
of Australia for operation as a National Facility managed by CSIRO. 
\bibliography{modrefs,psrrefs}
\bibliographystyle{mn}
\end{document}